\documentclass{emulateapj} %apt-get update; apt-get install emulateapj
\usepackage{apjfonts}

\newcommand{\mum}{\ifmmode{\rm \mu m}\else{$\mu$m}\fi}

\shorttitle{Unusual dust emission from PNe in the MCs}
\shortauthors{Bernard-Salas et al.}

\begin{document}

\title{Unusual dust emission from planetary nebulae in the Magellanic Clouds}

\author{J.~Bernard-Salas\altaffilmark{1},
E.~Peeters\altaffilmark{2,3},
G.C.~Sloan\altaffilmark{1}, 
S.~Gutenkunst\altaffilmark{1},
M.~Matsuura\altaffilmark{4,5},
A.G.G.M.~Tielens\altaffilmark{6},
A.A.~Zijlstra\altaffilmark{7},
J.R.~Houck\altaffilmark{1}
}

%\author{J. Bernard-Salas\altaffilmark{1}}
%\author{E. Peeters\altaffilmark{2}}
%\author{G.C. Sloan\altaffilmark{1}}
%\author{S. Gutenkunst\altaffilmark{1}}
%\author{P.W. Morris\altaffilmark{3}}
%\author{J.R. Houck\altaffilmark{1}}
%\author{et al.}

%\email{jbs@isc.astro.cornell.edu}

\altaffiltext{1}{Center for Radiophysics and Space Research, Cornell
University, 222 Space Sciences Building, Ithaca, NY 14853-6801, USA.}
\altaffiltext{2}{Department of Physics and Astronomy, The University
  of Western Ontario, London ON N6A 3K7, Canada.}
\altaffiltext{3}{SETI Institute, 515 N. Whisman Drive, Mountain View,
  CA 94043, USA.}
\altaffiltext{4}{National Astronomical Observatory of Japan, Osawa
  2-21-1, Mikata, Tokyo 181-8588, Japan.}
\altaffiltext{5}{Department of Physics and Astronomy, University
  College London, Gower Street, London WC1E 6BT, UK}
\altaffiltext{6}{NASA Ames Research Center, MS 245-3, Moffett Field,
  CA 94035, USA}
\altaffiltext{7}{Jodrell Bank Centre for Astrophysics, The University
  of Manchester, Manchester M13 9PL, UK}

\begin{abstract}

  We present a {\em Spitzer Space Telescope} spectroscopic study of a
  sample of 25 planetary nebulae in the Magellanic Clouds. The
  low-resolution modules are used to analyze the dust features present
  in the infrared spectra. This study complements a previous work by
  the same authors where the same sample was analyzed in terms of neon
  and sulfur abundances.  Over half of the objects (14) show emission
  of polycyclic aromatic hydrocarbons, typical of carbon-rich dust
  environments. We compare the hydrocarbon emission in our objects to
  those of Galactic H\,II regions and planetary nebulae, and LMC/SMC
  H\,II regions. Amorphous silicates are seen in just two objects,
  enforcing the now well-known-fact that oxygen-rich dust is less
  common at low metallicities. Besides these common features, some
  planetary nebulae show very unusual dust.  Nine objects show a
  strong silicon carbide feature at 11~$\mu$m and twelve of them show
  magnesium sulfide emission starting at 25~$\mu$m. The high
  percentage of spectra with silicon carbide in the Magellanic Clouds
  is not common.
  %About nine of the objects display a weak excess
  %between 16 and 22~$\mu$m.  
  Two objects show a broad band which may be attributed to hydrogenated
  amorphous carbon and weak low-excitation atomic lines.  It is likely
  that these nebulae are very young.  The spectra of the remaining
  eight nebulae are dominated by the emission of fine-structure lines
  with a weak continuum due to thermal emission of dust, although in a
  few cases the S/N in the spectra is low, and weak dust features may
  not have been detected.

%If these nebulae
%  have a disk, then probably in low metallicity nebulae this disk is
%  carbon-rich, together with a reservoir of silicon, which then
%  enhances the probability of forming this feature.

\end{abstract}


% Discussion on SiC, hardness of ISRF
% SiC evolution (age, processing pahs)

\section{INTRODUCTION}

Dust forms at the end of the life of stars, during the Asymptotic
Giant Branch (AGB) phase \citep{geh89}. Stars of low- and
intermediate-mass progenitors experience strong mass loss during the
AGB. For carbon-stars the radiation pressure on dust grains drives an
outflow of gas which is coupled to and triggers the formation of
molecules in a circumstellar envelope \citep[e.g.][]{hab96}. The
mechanism driving the mass-loss in oxygen-rich stars is still
debatable. \citet{woi} argued that the outflow in oxygen-rich stars
cannot be driven by pulsations given the low near-infrared opacity of
silicate grains, but he assumed models with only small amounts of iron
in the grains. Adding iron increases the opacity, and iron is a
significant component of inter-planetary dust particles and may be
present in AGB shells \citep{ngu08}. Whatever the details of the
envelope ejecta, the remaining envelope contracts onto a degenerate
C-O core, and as the effective temperature rises the ultraviolet
photons start to dissociate the molecules previously ejected
\citep{ber05}. At some point ionization of the ejected gas will occur
giving birth to the planetary nebula phase. A significant mass of the
envelope will not be photo-dissociated, where molecules and dust will
survive.  These neutral regions in the interstellar medium where the
chemistry and heating are governed by far ultraviolet photons are
referred to as photo-dissociation regions \citep[PDRs,][]{hol99}.

The peak of dust emission in planetary nebulae (PNe) occurs in the
mid-infrared ($\sim$20-40$\mu$m).  The infrared spectra of these
objects usually show dust features of either carbonaceous or silicate
material. If the nebula is carbon-rich (C/O$>$1), then most of the
oxygen is locked into CO. This is a very stable molecule and the
excess of carbon can form polycyclic aromatic hydrocarbons (PAHs),
common in the spectra of a great variety of astronomical environments.
If the nebula is oxygen rich, then the excess of oxygen can form
amorphous and/or crystalline oxides and silicates.

A number of galactic PNe show spectral evidence for both carbonaceous
and silicate dust. These include planetary nebulae with WC central
stars such as BD$+$30\,3639, CPD$-$56\,8032 and He~2-113
\citep{coh02,ber05} as well as PNe resulting from very massive
progenitors (like NGC~6302). In addition, some peculiar post-AGB
objects (e.g. the Red Rectangle) also show evidence for mixed
chemistry.  Often the components are spatially separated with the
silicates in a long-lived disk \citep{wat98} and the carbonaceous dust
in the outflow \citep{llo90,coh99}. Observation of this dual chemistry
in PNe spectra is independent of their gas phase chemical abundances
within the PNe.  The ratio of nebulae with double chemistry is higher
in the Bulge as shown by \citet{gut08} where out of 11 objects, six
display both PAH and silicates features, and it is probably linked to
their binary evolution or to a late phase of carbon-rich mass loss. In
particular, a long lived stable disk may trap early (O-rich) ejecta
through a subsequent C-rich mass loss phase in such binary systems.

Our general understanding of how these dust features form and evolve
is still poor. For instance, PAH bands are known to vary not only in
strength but also in profile; however the cause of these variations is
still not known. Some hypotheses have included effects of metallicity,
radiation field, grain size, grain charge, and dust composition
\citep{pee02,die04,slo05,boe08, bau08,bau09,job08,cam09}. The PAHs in
PNe show in fact a larger variation from object to object than in
other astronomical environments \citep{pee02}.

The Infrared Space Observatory ({\em ISO}) made possible the detailed
study of the infrared spectra of many PNe in the Milky Way. However,
due to the lack of sensitivity, {\em ISO} could not observe
spectroscopically PNe outside the Galaxy. The Large and Small
Magellanic Clouds (LMC and SMC, respectively) offer a great
opportunity to study the dust in PNe in a different (metal poor)
environment. The {\em Spitzer Space Telescope} (SST) with its enhanced
sensitivity is more suited to study dust in the Magellanic Clouds
(MCs). In a recent paper, \citet{ber08} presented the high-resolution
10-37~$\mu$m spectra of a sample of 26 PNe in the MCs.  In this paper,
we present the results of the low-resolution modules on board the {\em
  SST} which are more suited to the study of dust. These results are
compared to those found in PNe in the Galaxy and to H\,II regions in
an attempt to characterize the kind of dust present in the MC PNe and
how it compares to the dust in the Galaxy.

Using {\em IRAS} data, \citet{len89} studied the properties of the
dust of several hundreds of PNe. They found that the grain size and
total dust mass decreases as the nebula evolves, which they attributed
to sputtering and grain-grain collisions. \citet{sta99} have argued,
however, that this results from an artifact in their method of
analysis.  Recently, \citet{sta07} used {\em Spitzer} spectra of a
sample of 41 MC PNe to correlate the dust features with the physical
parameters of the central stars and the PNe morphology. They find that
the temperature of the dust decreases as the nebula evolves and that
the production of dust decreases at lower metallicity when compared to
Galactic PNe. In this paper, we focus on the characterization of the
dust features, their dependence on the environment as given by the
strength of radiation field, and how the features compare to H\,II
regions and Galactic PNe.

The paper is organized as follows. The next section explains the
observations and data reduction. In \S3, we explore and analyze the
hydrocarbon features. Two carbonaceous dust features which are
prominent in our spectra, silicon carbide (SiC) and magnesium sulfide
(MgS) are presented in \S4 and \S5 respectively.  A discussion on the
silicate features is given in \S6. Other features present in the IRS
data are discussed in \S7. Finally the conclusions are presented in
the last section.

\section{OBSERVATIONS AND DATA REDUCTION}

The observations were taken with the {\it Spitzer Space Telescope}
\citep{wer} and were part of the IRS Guaranteed Time Observations
(GTO) program (ID 103). From this program, SMP~LMC~11 is not included
in this sample as its spectrum was already published in \citet{ber06}
and found to be a pre-planetary nebula.  The observations consisted of
high- and low-resolution spectra of 25 PNe taken with the Infrared
Spectrograph \citep[IRS\footnote{The IRS was a collaborative venture
  between Cornell University and Ball Aerospace Corporation funded by
  NASA through the Jet Propulsion Laboratory and the Ames Research
  Center.},][]{hou}. As mentioned in the introduction we refer only to
the results using the low-resolution modules (SL and LL). These
modules cover a wavelength region from 5.5 to 37~$\mu$m with a
resolution of 60-120.  Details on the assumed coordinates and pointing
strategy are given by \citet{ber08}.  The data were taken using the
IRS Staring Mode observing template which produces spectra in two {\em
  nod} positions (at 1/3 and 2/3 of the slit length). The spectral
images were processed with the S13.2 version of the {\em Spitzer}
Science Center's pipeline and using a script version of {\it Smart}
\citep{hig}. The data reduction started from the {\it droop} files
which are similar to the most commonly used {\it bcd} data and only
lack the flatfield and stray-cross-light removal.  The rogue pixels
were removed using the {\it irsclean} tool which is available from the
SSC website\footnote{http://ssc.spitzer.caltech.edu}.  When different
cycles were present for a given observation these were combined to
improve the S/N. This was done for each module and {\em nod}
separately. Then, for a given module the two {\em nod} positions were
differentiated to remove the contribution from the background (e.g.
SL1 nod1 minus SL1 nod2 and vice-versa). Spectra were extracted from
the resultant 2D-images using variable column extraction, which is
fixed to four pixels at the wavelength center of each module.  The
calibration was performed by dividing the resultant spectrum by that
of the calibration star (which was extracted in the same way as the
target), then multiplying by its template \citep[][ and Sloan et al.
in prep]{coh}. The calibration stars used are HR\,6348 for the SL
module and HD\,173511 for LL.  Remaining glitches which were not
present in both {\em nod} positions or in the overlapping region
between orders were removed manually. We note that there is another
source close to SMP-SMC11 in the SL slit, and may contaminate to some
extent the PAH fluxes (between 10-40\%) quoted in Table~1 for this
source.

In some of the spectra a discontinuity is present between the
different modules ($\sim$10-30\% in flux). These discontinuities
result from slight mispointings.  We used coordinates given by
\citet{sta02, sta03} and \citet{lei97} and performed Peak-Up
acquisition which produces 0.4\arcsec~accuracy. However, the SL slit
is very narrow (3.6\arcsec) and flux losses can occur if the
coordinates are not very accurate. In fact, there are differences of
1-2\arcsec~between the coordinates given by \citet{sta02, sta03} and
\citet{lei97} when they have objects in common in their sample. The LL
slit is 10.5\arcsec~across, wide enough to account for pointing
uncertainties, and thus the rest of the spectra were scaled to match
this module.

Figure~1 shows the extracted spectra, where the most prominent lines
and features are marked in the top of the figure\footnote{SMP~LMC~83
  is not shown in the Figure because its spectrum was already
  presented in \citet{ber04}.}.  PAHs are present in most of the
spectra, and two objects show amorphous silicates. No object is seen
with mixed chemistry. Eight objects show no apparent PAHs or silicates
and just a very weak continuum with strong forbidden lines. Table~1
gives a summary of the main features seen in the spectra and indicates
the high-excitation PNe and those with Wolf-Rayet features.

\begin{figure*}
  \begin{center}
  \includegraphics[width=16cm]{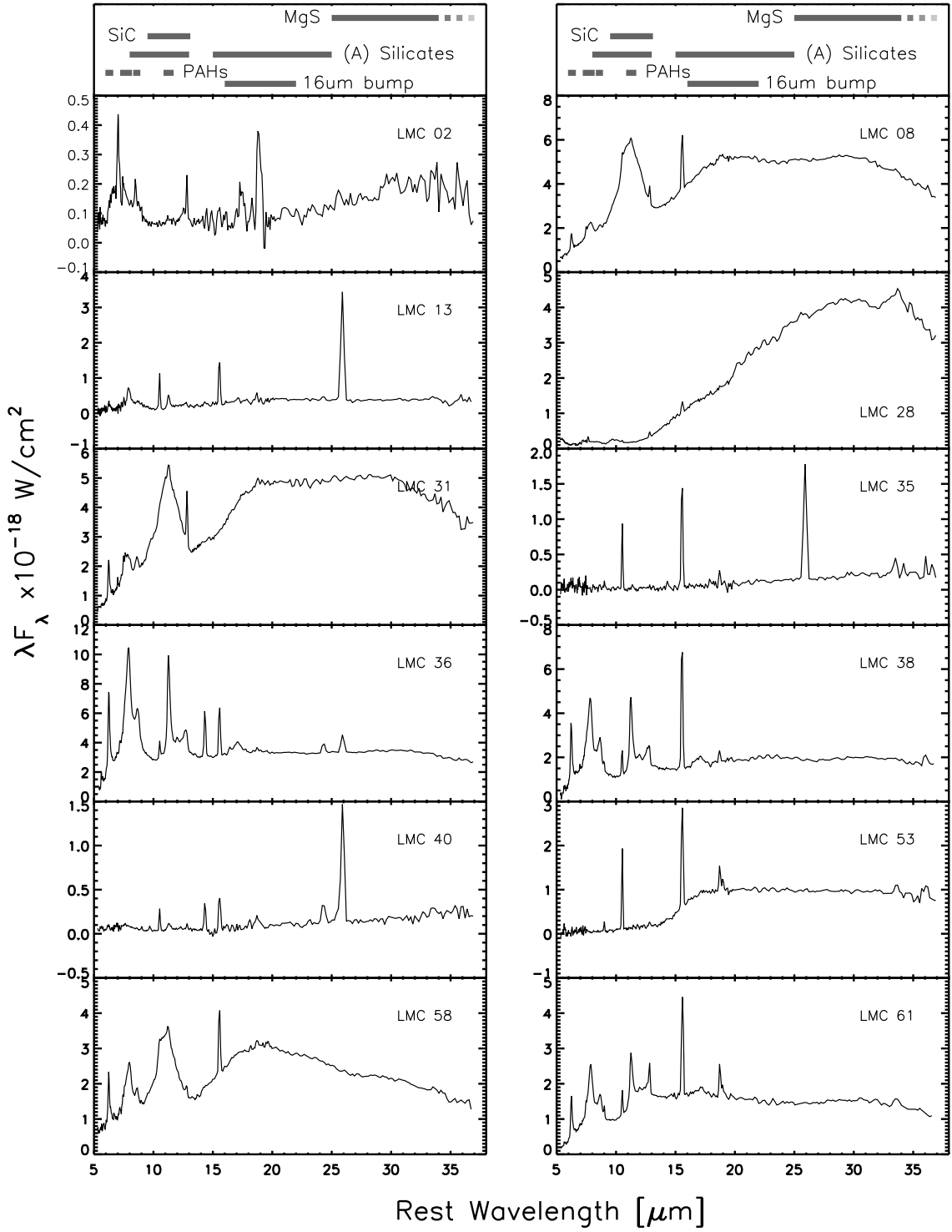}
  \end{center}
  \caption{{\em Spitzer}-IRS 's low resolution spectra of the MC PNe.
    The most prominent bands are given in the legend (top
    panels).\label{spectra_f}}
\end{figure*}

\setcounter{figure}{0}

\begin{figure*}
  \begin{center}
  \includegraphics[width=16cm]{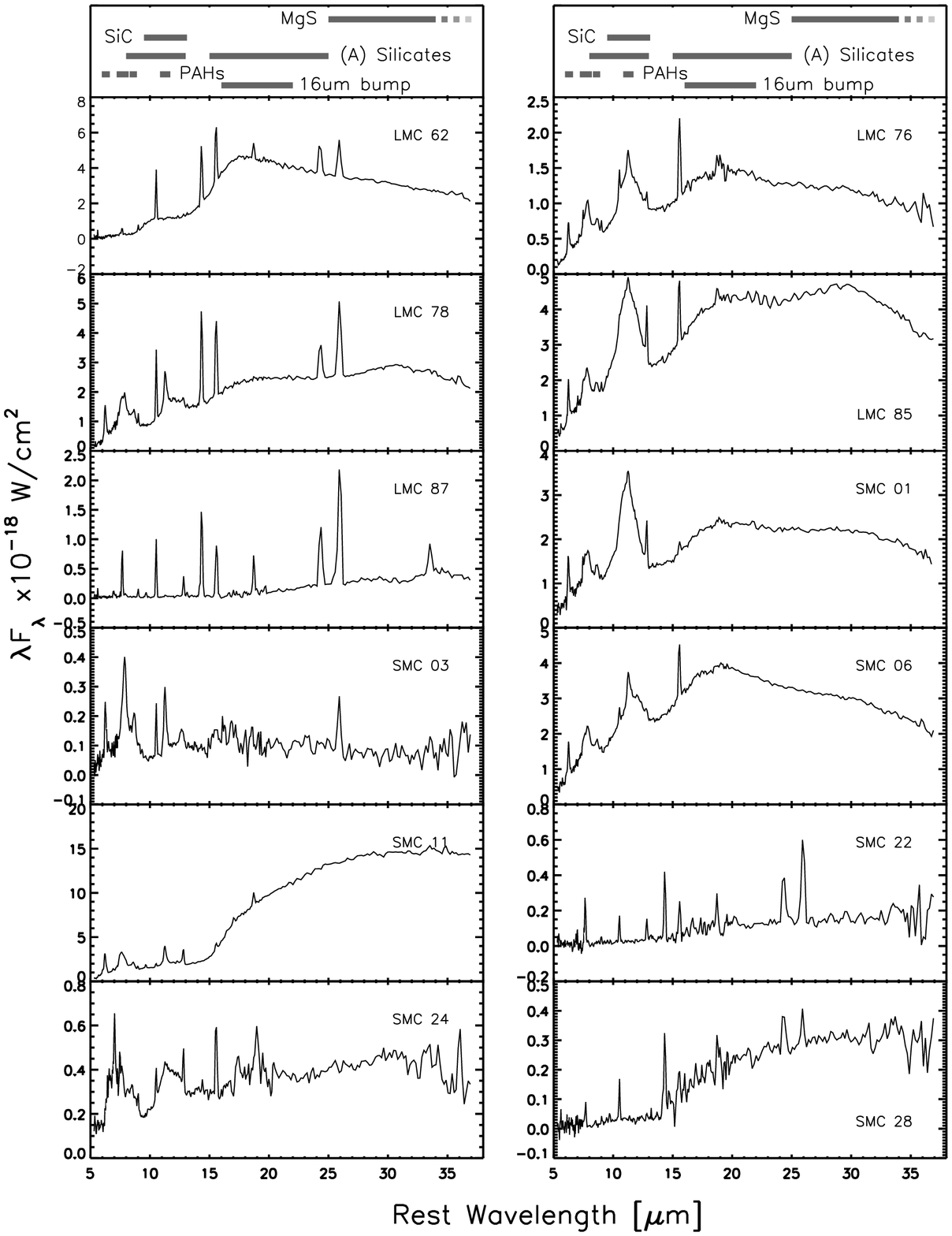}
  \end{center}
  \caption{Continuation Fig.~1.\label{spectra_f}}
\end{figure*}

\tabletypesize{\scriptsize}

\begin{deluxetable}{c c c c c c c c}
  %\tablecolumns{7}
  \tablewidth{0pc}
  \tablecaption{Sources classification.\label{classification_t}}

  \tablehead{\colhead{Object} &
  \colhead{HE\tablenotemark{a}} & \colhead{WR} & \colhead{PAHs\tablenotemark{b}} & 
  \colhead{SiC} & \colhead{MgS} & \colhead{Silicates} 
  & Other Feat.\tablenotemark{c}}

  \startdata    

  SMP LMC~02  & ... & ... & ... & ... & ... & ... & HAC       \\
  SMP LMC~08  & ... & ... & B,B & Y   & Y   & ... & Bump16     \\
%  SMP LMC~11\tablenotemark{d}  & ... & ... & ... & ... & ... & ... & Mol. Abs.   \\
  SMP LMC~13  &   Y & ... & B,B & ... & ... & ... & ...   \\
  SMP LMC~28  &   Y & ... & ... & ... & ... & Y?  & ...   \\
  SMP LMC~31  & ... & Y?  & B,A & Y   & Y   & ... & Bump16    \\
  SMP LMC~35  &   Y & ... & ... & ... & ... & ... & ...   \\
  SMP LMC~36  &   Y & ... & B,B & ... & Y   & ... & Plateau15 \\
  SMP LMC~38  & ... & Y   & A,B & ... & Y   & ... & Plateau15     \\
  SMP LMC~40  &   Y & ... & ... & ... & ... & ... & ...   \\
  SMP LMC~53  & ... & ... & ... & ... & ... & Y   & ...   \\
  SMP LMC~58  & ... & Y   & B,B & Y   & Y   & ... & Bump16    \\
  SMP LMC~61  & ... & Y   & B,B & ... & Y   & ... & Plateau15     \\
  SMP LMC~62  &   Y & ... & ... & ... & ... & Y   & ...   \\
  SMP LMC~76  & ... & ... & B,B & Y   & Y   & ... & Bump16    \\
  SMP LMC~78  &   Y & ... & AB,B& Y   & Y   & ... & Bump16    \\
  SMP LMC~83\tablenotemark{d}  &   Y & Y\tablenotemark{e}    & ... & ...  & ... & ...  & ...   \\
  SMP LMC~85  & ... & ... & B,B & Y   & Y   & ... & Bump16 \\
  SMP LMC~87  & ... & ... & ... & ... & ... & ... & ...   \\
  &      &    &     &  &   &    \\
  SMP SMC~01  & ... & ... & AB,AB& Y  & Y   & ... & Bump16    \\
  SMP SMC~03  & ... & ... & B,B & ... & ... & ... & ...   \\
  SMP SMC~06  & ... & Y   & B,B & Y   & Y   & ... & Bump16    \\
  SMP SMC~11  & ... & ... & A,A & ... & ... & ... & ...   \\
  SMP SMC~22  &   Y & ... & ... & ... & ... & ... & ...   \\
  SMP SMC~24  & ... & ... & ?   & Y   & Y   & ... & HAC    \\
  SMP SMC~28  &   Y & ... & ... & ... & ... & ... & ...   \\

  \enddata

  \tablenotetext{a}{High Excitation PNe, those with \ion{[O}{4]} and
    \ion{[Ne}{5]} lines.}
  \tablenotetext{b}{PAH classification of the 6.2 band and 7.7~$\mu$m
    complex respectively according to the scheme by \citet{pee02}.}
  \tablenotetext{c}{Other features. The Bump16 refers to the broad
    excess between 16-22~$\mu$m; Mol.Abs. indicates molecular
    absorption; Plateau15 refers to the underlying PAH plateau due to
    C-C-C modes at 15.5-20~$\mu$m; and HAC is the acronym for
    Hydrogenated Amorphous Carbon.}
  \tablenotetext{d}{This object was presented in
    \citet{ber04}. It was part of the original program but was observed
    during In Orbit Checkout.}
  \tablenotetext{e}{Instead of the usual [WC], SMP~LMC~83 has a [WN]
    central star.}
\end{deluxetable}

  \tabletypesize{\normalsize}

\section{HYDROCARBONS}

\subsection{PAHs}

Out of the PNe listed in Table~1, 14 of them show PAH features. In
these objects, the most typical features at 6.2~$\mu$m, 7.7~$\mu$m,
8.6~$\mu$m, and 11.2~$\mu$m are always present. The 12.7~$\mu$m
feature is detected in five of the PNe and the 16.4~$\mu$m band and
17~$\mu$m complex are seen in just three of the nebulae.  The
11.2~$\mu$m band is in some cases on top of a broad SiC feature (see
\S4). This same SiC feature makes it difficult to detect the
12.7~$\mu$m complex in some of the objects. The small PAH bands at 6.0
and 11.0$\mu$m are also seen.

%In addition to these 14
%PNe, the spectrum of SMP~LMC~11 shows absorption features
%corresponding to acetylene, benzene and other molecules which are the
%precursors of the PAH features \citep{ber06}.

  \subsubsection{The Strength of the PAHs}

  The strengths of the PAH emission features were determined by
  integrating the flux of the feature above an adopted continuum in
  the nod-combined spectra and are given in Table~2. The baseline for
  the continuum was derived by fitting a spline function to selected
  regions of the spectra around each PAH. The integration limits were
  set to the following values: 6.1-6.6~$\mu$m for the 6.2~$\mu$m PAH
  band, 7.20-8.3~$\mu$m for the 7.7~$\mu$m PAH, 8.3-8.88~$\mu$m for
  the 8.6~$\mu$m PAH, and 11.1-11.7~$\mu$m for the 11.2 feature. This
  method produced very solid results which were reproducible to within
  5\% for most objects. In some cases the 11.2$\mu$m PAH feature sits
  on top of a much broader and stronger SiC feature. To make sure that
  the continuum underlying this feature is well determined we
  over-plotted the PAH profile of the Orion nebula when measuring the
  feature. In addition we measured this feature using the
  high-resolution (HR) spectra presented in \citet{ber08}, where the
  features are better disentangled (but the spectra are more affected
  by rogue pixels), and found that the measured values in most of the
  cases were well within 10\% (Table~2).

  \begin{figure}
    \begin{center}
      \includegraphics[width=8.5cm]{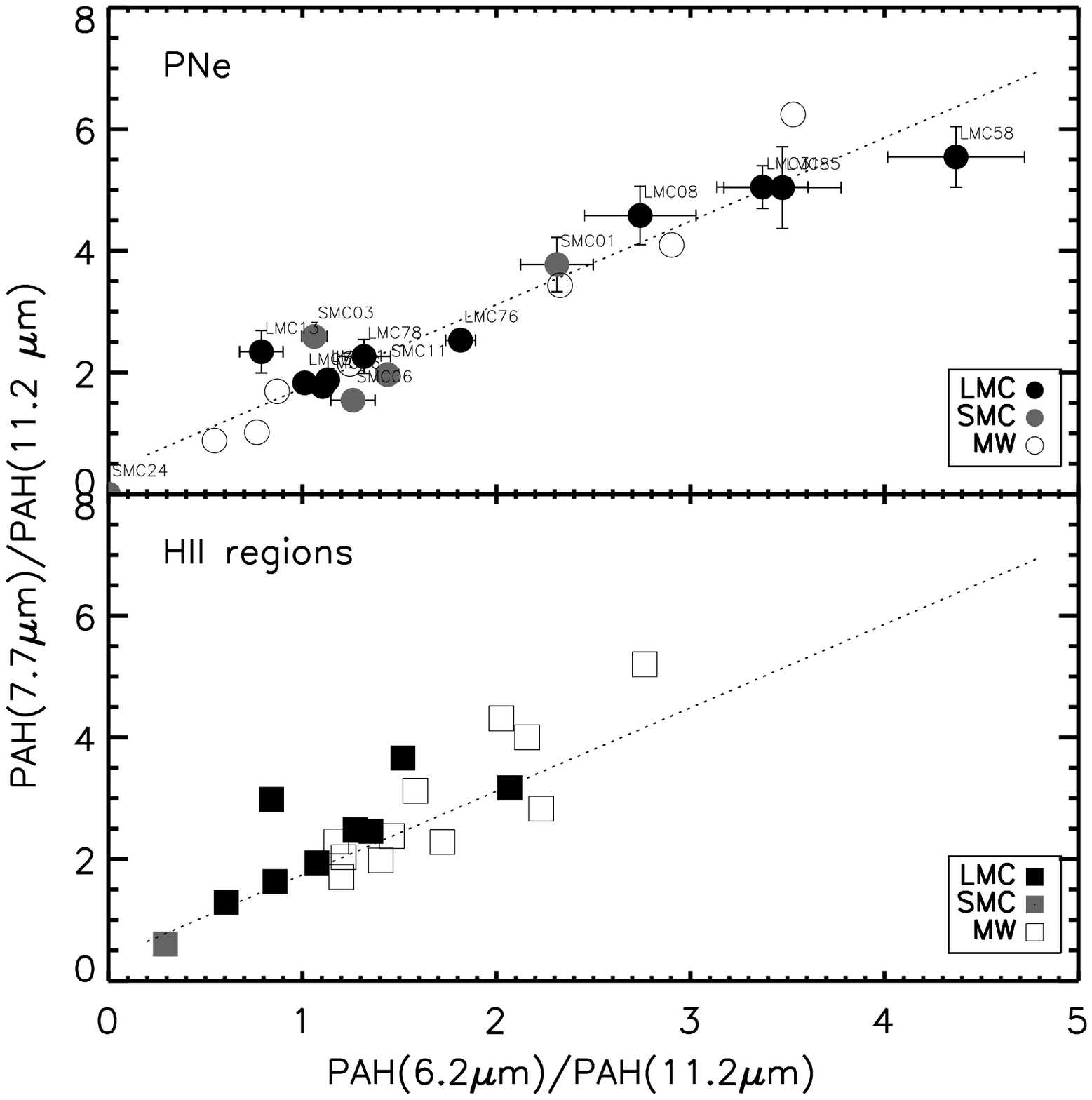}
    \end{center}
    \caption{The ratio of ionized and neutral PAHs in PNe, compared
      with that of H\,II regions from the sample of
      \citet{ver}.\label{ratiopah_f}}
  \end{figure}

  An important characteristic of the PAHs is that their emission is
  influenced by the degree of ionization in the PDR
  \citep{lan96,all99}. A mixture dominated by neutral PAHs will emit
  strongly in the 11.2~$\mu$m band, while the bands at 6.2~$\mu$m,
  7.7~$\mu$m, and 8.6~$\mu$m are stronger in ionized PAHs. In Figure~2
  the ratio of the 7.7$\mu$m/11.2$\mu$m PAH bands (ionized/neutral) is
  plotted against the 6.2$\mu$m/11.2$\mu$m (ionized/neutral) ratio for
  the MC PNe together with a representative sample of ratios in
  Galactic PNe, and compared with ratios found by \citet{ver} for
  H\,II regions.  Higher ratios indicate a higher degree of ionisation
  because the 7.7 and 6.2~$\mu$m PAHs are stronger in ionized PAHs
  than the 11.2~$\mu$m PAH, and it can be seen that both ratios
  correlate well with each other.  The correlation in the PNe has a
  slope of 1.3$\pm$0.1, which agrees within uncertainties with the
  slope derived for the H\,II regions (1.5$\pm$0.2).  \citet{ver}
  hinted that the H\,II regions seem to follow a relation with
  metallicity; the MW H\,II regions having the largest ratios while
  LMC H\,II regions have somewhat lower values and the only SMC H\,II
  region in their sample has the lowest ratio.  While this relation
  was hampered by the low number of objects (especially in the SMC)
  recently \citet{leb08} in their study of the 3 giant H\,II regions
  NGC3603 (in the MW), 30~Doradus (LMC) and NGC~346 (SMC) find that
  this metallicity relation holds for these regions as well. All other
  things being equal, when the metallicity decreases, the electron
  density in the PDR (due to C ionization) will decrease. This results
  in a decreased neutralization rate and hence a shift to a higher
  degree of ionization, which is contrary to the observations of the
  H\,II regions.
  % This is intriguing because one would expect that at lower
  % metallicities the hardness of the radiation field is higher and
  % ionisation should increase; however the degree of ionisation not
  % only depends on the incident radiation field but also on the
  % density and temperature of the PDR.
  The PNe do not show such a segregation in metallicity, indeed the
  PNe in the MW, LMC and SMC are scattered all across Figure~2.  This
  is not surprising considering that PNe are not a homogeneous sample;
  they all have different ages.

  The five PNe with the highest PAH degree of ionisation (Fig.~2) are
  those with the highest SiC fluxes (SMP~LMC~08, SMP~LMC~31, SMP~LMC~58,
  SMP~LMC~85, and SMP~SMC~01). This is not due to uncertainties in the
  determination of the PAH 11.2~$\mu$m flux above the SiC feature
  because the measured PAH flux in the high-resolution spectra (where
  both the PAH and SiC can be clearly separated) yields the same flux
  as in the low-resolution measurements.

\tabletypesize{\scriptsize}

  \begin{deluxetable*}{c c c c c c c c c c c c c c c}
  %\begin{deluxetable*}{c c c c c c c c c c c c c c c}
  \tablewidth{0pc}
  \tablecaption{PAH and SiC fluxes.\tablenotemark{a}\label{pahflux_t}}

  \tablehead{\colhead{Object} & \multicolumn{2}{c}{6.2$\mu$m} &
    \multicolumn{2}{c}{7.7$\mu$m} & \multicolumn{2}{c}{8.6$\mu$m} &
    \multicolumn{2}{c}{11.2$\mu$m} & \multicolumn{2}{c}{11.2$\mu$m
      (HR)\tablenotemark{b}} & \multicolumn{2}{c}{SiC}  & \multicolumn{2}{c}{SiC}\\
    & \colhead{Flux} & \colhead{Error} & \colhead{Flux} &
    \colhead{Error}  & \colhead{Flux}  & \colhead{Error} &
    \colhead{Flux}   & \colhead{Error} & \colhead{Flux} &
    \colhead{Error}  & \colhead{Flux}  & \colhead{Error} & 
    \colhead{EW\tablenotemark{c}} & \colhead{Error}}

  \startdata    

LMC-SMP08  &   2.22  & 0.08 &   3.71 &  0.13 &  0.63 &  0.04 &  0.81 &  0.08 &  1.18 &  0.03 &  55.98 & 0.35 &  23.21 & 0.32 \\
LMC-SMP13  &   0.67  & 0.09 &   1.99 &  0.28 &  0.09 &  0.09 &  0.85 &  0.04 &  0.91 &  0.01 &  ...   & ...  &  ...   & ...  \\
LMC-SMP31  &   3.54  & 0.18 &   5.30 &  0.27 &  1.28 &  0.06 &  1.05 &  0.05 &  1.25 &  0.06 &  46.90 & 0.09 &  22.31 & 0.04 \\
LMC-SMP36  &  16.01  & 0.56 &  25.63 &  0.36 &  3.70 &  0.14 & 14.50 &  0.16 & 14.03 &  0.12 &  ...   & ...  &  ...   & ...  \\
LMC-SMP38  &   6.94  & 0.13 &  12.55 &  0.31 &  2.67 &  0.22 &  6.86 &  0.11 &  5.45 &  0.04 &  ...   & ...  &  ...   & ...  \\
LMC-SMP58  &   4.50  & 0.10 &   5.71 &  0.26 &  0.82 &  0.12 &  1.03 &  0.08 &  1.07 &  0.03 &  29.72 & 0.31 &  21.40 & 0.28 \\
LMC-SMP61  &   3.65  & 0.05 &   6.07 &  0.17 &  1.00 &  0.05 &  3.23 &  0.10 &  3.23 &  0.05 &  ...   & ...  &  ...   & ...  \\
LMC-SMP76  &   1.47  & 0.03 &   2.05 &  0.12 &  0.41 &  0.05 &  0.81 &  0.03 &  0.87 &  0.02 &  11.43 & 0.21 &  17.19 & 0.35 \\
LMC-SMP78  &   3.49  & 0.12 &   6.00 &  0.45 &  0.85 &  0.23 &  2.65 &  0.26 &  2.93 &  0.02 &   8.52 & 1.27 &  8.75  & 1.48 \\
LMC-SMP85  &   2.71  & 0.11 &   3.93 &  0.43 &  0.71 &  0.10 &  0.78 &  0.06 &  0.86 &  0.06 &  45.12 & 0.43 &  24.65 & 0.40 \\
           &         &      &        &       &       &       &       &       &       &       &        &      &        &      \\
SMC-SMP01  &   2.15  & 0.13 &   3.51 &  0.37 &  0.75 &  0.07 &  0.93 &  0.05 &  0.97 &  0.03 &  30.39 & 0.28 &  26.84 & 0.51 \\
SMC-SMP03  &   0.52  & 0.03 &   1.27 &  0.04 &  0.23 &  0.02 &  0.49 &  0.01 &  0.51 &  0.02 &  ...   & ...  &  ...   & ...  \\
SMC-SMP06  &   2.95  & 0.14 &   3.61 &  0.21 &  0.38 &  0.18 &  2.34 &  0.18 &  2.12 &  0.04 &  18.42 & 0.84 &  10.61 & 0.55 \\
SMC-SMP11\tablenotemark{d}  &   6.43  & 0.12 &   8.78 &  0.35 &  1.39 &  0.10 &  4.47 &  0.16 &  4.92 &  0.11 &  ...   & ...  &  ...   & ...  \\
SMC-SMP24  &   ...   & ...  &   ...  &  ...  &  ...  &  ...  &  0.15 &  0.03 &  ...  &  ...  &   3.47 & 0.17 &  19.88 & 1.28 \\ 

  \enddata

  \tablenotetext{a}{Fluxes in units of 10$^{-16}$ W~m$^{-2}$}
  \tablenotetext{b}{Measured in the high-resolution spectra.}
  \tablenotetext{c}{Equivalent width is given in $\mu$m.}
  \tablenotetext{d}{PAH fluxes for this source may be
      contaminated by the presence of another source nearby (see
      \S2).}
  %\tablenotetext{c}{The contribution of the Pf$\alpha$ line at
  %  7.45~$\mu$m has been removed from the 7.7~$\mu$m PAH flux.}
  \end{deluxetable*}
  %\end{deluxetable*}

  \tabletypesize{\normalsize}

  \cite{gor08} studied the relation of the PAH features with the
  hardness of the radiation field. They find a trend of decreasing PAH
  equivalent width (EW) with hardness of radiation in M101.  This
  correlation leads them to argue that the variations in the strength
  of the PAH features are mainly the effect of processing of the dust
  grains. While not shown, our sample does not show any clear trend of
  PAH EW with \ion{[Ne}{3]}/\ion{[Ne}{2]} ratio (or the sulfur ratio)
  with the points scattered along the plot.  However, in Figure~1, the
  presence of PAHs in the spectra is anti-correlated with the presence
  of the \ion{[O}{4]} and \ion{[Ne}{5]} lines (see Table~1). Surely
  PAHs will not survive the hard radiation field needed to excite
  these ions, but because we are integrating over the whole nebula one
  could expect emission from PAHs in the photo-dissociation region (in
  the neighborhood of HII regions). Some Galactic PNe with high
  excitation lines also show PAHs (e.g.  NGC~7027). Like in our
  sample, the larger sample studied by \citet{sta07} also shows this
  trend, although a few objects with high excitation lines show very
  weak PAHs. It is possible then, that if the radiation field is high
  enough in the LMC and SMC, the PAHs are destroyed. The fact that
  this seems not to apply to the Galactic PNe may be ascribed to a
  size distribution.
  %In the LMC and especially in the SMC the lower
  %metallicity may prevent the formation of bigger PAHs which can be
  %more easily destroyed than in the Milky Way. 
  %We note that PAHs in
  %H\,II regions seem to grow bigger with decreasing metallicity
  %(Lebouteiller et al.  2006, Bernard-Salas et al. in prep), but there
  %is no indication that the formation process in both regions (PNe and
  %H\,II regions) has to be the same.

  \subsubsection{Profiles}

  It is well established that the profiles of the main PAH bands vary
  from object to object and spatially within extended objects
  \citep[cf.][and references therein]{pee04}. \citet{pee02} and
  \citet{die04} proposed a classification of the profiles of the main
  PAH bands based upon the {\em ISO} spectra of a variety of Galactic
  objects.  In these studies, most PNe (all but five) belonged to
  class B, with the 6.2~$\mu$m PAH peaking between 6.24-6.28~$\mu$m,
  the 7.7~$\mu$m PAH between $\sim$ 7.8-8.0~$\mu$m and a red-shifted
  8.6~$\mu$m PAH band. The three exceptions are Hb~5, IRAS~21282+5050,
  and NGC~7027, which show a mixing in classes (e.g. class A for one
  feature and B for others). The latter three sources together with
  BD+30~3639 also have an 11.2~$\mu$m PAH profile belonging to class
  A(B).

  \begin{figure*}
    \begin{center}
      \includegraphics[width=18cm]{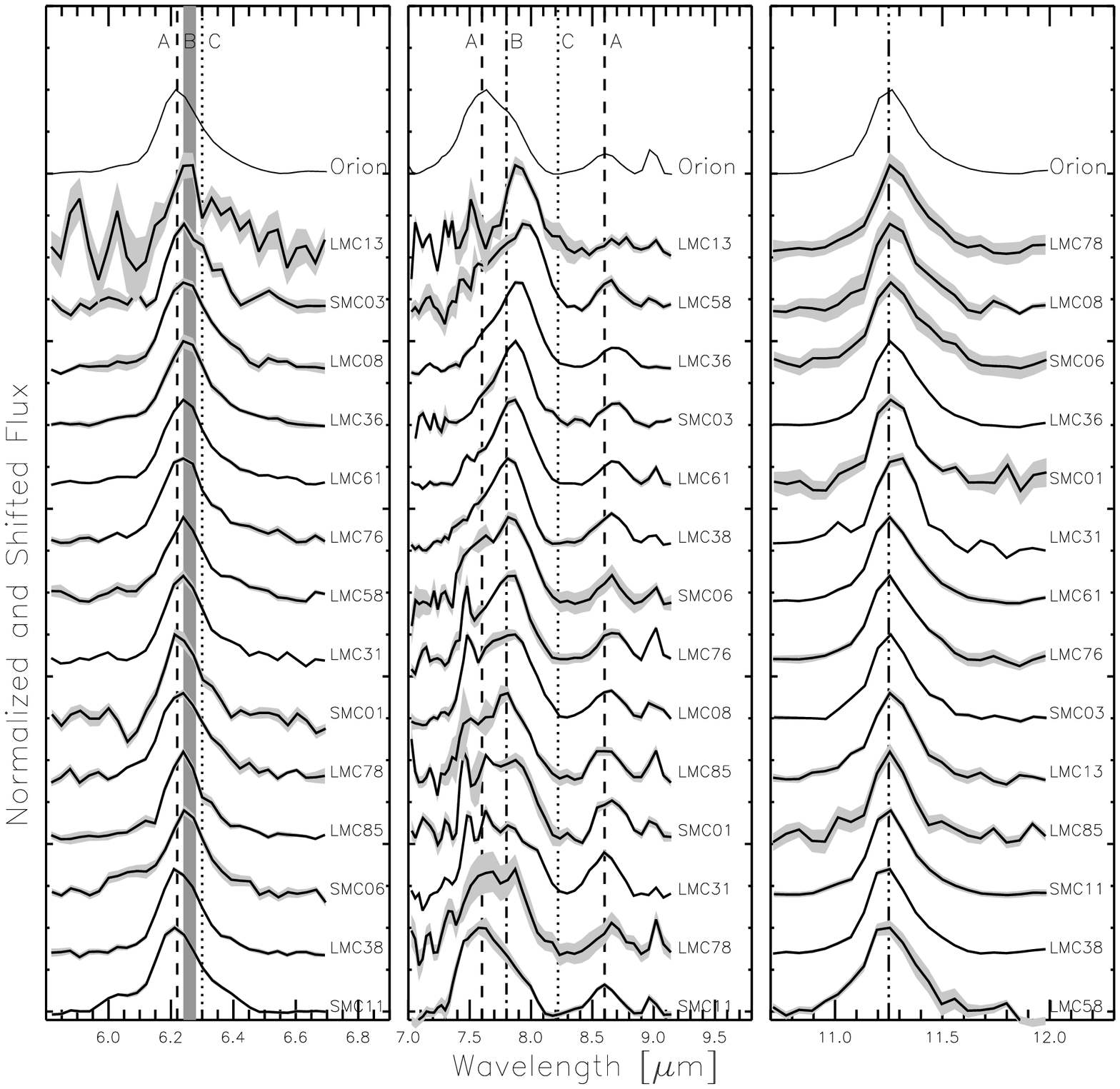}
    \end{center}
    \caption{PAH profiles (in colour) compared to that of Orion (thin
      black line). The profile classification given by \citet{pee02}
      and \citet{die04} is given on the top of the figure. The grey
      area around a given spectrum indicates the uncertainty in the
      flux.\label{profile_f}}
  \end{figure*}

  Figure~3 shows the continuum-subtracted spectra of the PNe ordered
  according to the center wavelength of the 6.2~$\mu$m PAH (left
  panel), the 7.7~$\mu$m PAH (middle) and 11.2~$\mu$m PAH (right).
  The center wavelength is taken as the position at which the total
  flux of the feature is half and is used to derive the equivalent
  width. The vertical lines and band indicate the nominal positions
  for the 3 classes in the \citet{pee02} and \citet{die04} scheme.  To
  guide the eye, the IRS spectra of Orion, which is a typical class A
  6.2~$\mu$m band, is over-plotted in black on top of each panel.
  Clearly, the low resolution of the SL and LL modules does not allow
  one to discern small variations (0.069~$\mu$m around 6-8~$\mu$m) as
  done in e.g.  \citet{pee02}, but it certainly can be used to study
  global trends, especially for the 7.7~$\mu$m complex. Hence the IRS
  is still highly suitable to investigate PAH profile variations
  \citep[e.g.][]{slo05,slo07}.  For instance, it can be seen from
  Figure~3 that the profile of the 6.2~$\mu$m PAH band shows variation
  in the onset of the blue wing and the peak position.
 %% can't decide between order according to peak or center wavelength
 %% for 6.2 peak seem to be better in this case.
  Three PNe seem to be in the class A regime and the rest are class B.
  Variations in the 7.7 $\mu$m complex are very obvious. The profile
  ranges from class A (lower two profiles) to class B. Note that while
  profiles of class A remain fairly similar, large variations are
  present within class B. The latter is actually defined as having a
  dominant ``7.8'' $\mu$m component of which the peak position and the
  entire profile can be red-shifted considerably \citep{pee02}.  This
  can also be seen from Figure~3.
  %% here I prefer the center wavelength order I think for the peak
  %% order, the lower two might be misplaced due to the atomic lines?
  Small variations in the peak position of the 8.6~$\mu$m profile seem
  to be apparent as well.  Unfortunately, due to the low spectral
  resolution no significant variations can be detected regarding the
  11.2~$\mu$m PAH band profile (Fig.~3). A possible exception could be
  SMP~LMC~31. In general, sources exhibiting a class B 6.2 profile
  belong to class B for the other profiles as well. However, note that
  some PNe belong to mixed classed (e.g. SMP~LMC~85).

  There have been several hypotheses to explain the changes in
  wavelength of the peak of the PAHs. N-substituted PAHs (PANHs;
  Peeters et al.  2002, Hudgins et al. 2005) can produce emission at
  6.2 $\mu$m. Indeed, pure PAHs exhibit emission longwards of 6.3
  $\mu$m in this specific region.  Hence, within this framework, both
  pure PAHs and PANHs are needed to explain the observed variations.
  Recently, \citet{bau08,bau09} found that negatively charged PAHs are
  important in determining the overall emission spectrum, however, the
  spectra of negatively charged PANHs were not considered in the
  original studies of PANHs.  Unlike for pure PAHs, the addition of an
  electron to PANHs redshifts the CC stretching mode, suggesting that
  charge variations in PANHs alone can explain the observed variations
  in the 6.2 PAH band \citep{bau09}.
    % If confirmed by further study, this suggests that the PAH
    % population present in space may well be comprised of
    % nitrogenated PAHs only \citep{bau09}.
  The amount of blue-shifting due to N-inclusion depends on the N/C
  ratio and the position of the N-atom inside the C skeleton
  \citep{hud05}. In a low metallicity environment it may be expected
  that less nitrogen compared to carbon is available (due the higher
  efficiency of the 3$^{rd}$ dredge-up at low metallicities).
  Therefore, the relative fraction of pure PAHs with respect to PANHs
  may be higher in the MC, and as a consequence, the PAH emission
  bands may be more shifted to the red (assuming that a mixture of
  pure PAHs and PANHs is needed).  However, this does not seem to be
  supported by the considered sample. While some sources have indeed a
  class A 6.2~$\mu$m profile, the observed variations in the MC PNe
  are also similar to those observed in our MW.  Certainly, it would
  be interesting to look at the C/N abundance ratio to investigate
  this more thoroughly, but reliable determinations of C and N
  abundances are lacking in our sample.

  The ratio of $^{13}$C/$^{12}$C has also been proposed to explain the
  observed variations in the 6.2~$\mu$m PAH band \citep{wad03}.
  However, while an enhanced $^{13}$C/$^{12}$C ratio does shift the
  emission towards other wavelengths, it will redshift the emission.
  Given that pure PAHs exhibit emission longwards of 6.3 $\mu$m in
  this region,  enhanced $^{13}$C/$^{12}$C in pure PAHs will not
  bring these bands towards 6.2 $\mu$m.

  Laboratory and theoretical PAH spectra reveal that small and large
  PAHs emit at different wavelengths in the 7.7 $\mu$m region
  \citep{bau08,bau09}. In particular, small PAHs exhibit strong
  emission near $\sim$7.6 $\mu$m, while large PAHs emit near $\sim$
  7.8 $\mu$m.  \citet{bau08,bau09} suggest that the observed
  difference between class A and class B sources is related to a
  variation in the size distribution of the PAH family.  In addition,
  these studies revealed changes in the specific position of the
  emission near 7.7$\mu$m with PAH charge. Large PAH anions emit at
  slightly longer wavelengths than PAH cations. Hence, variation in
  the peak position of the 7.8 $\mu$m component (and the 6.2 band, see
  above), and thus variation within the class B profiles, may be
  related to the variation in relative amount of large PAH cations and
  anions.  \citet{job08} invoked a new component in their mathematical
  decomposition of observed PAH spectra with redshifted bands at 7.9
  and 8.65$\mu$m. These authors found that PAH anions may be plausible
  candidates for this component. Observationally, this charge
  dependence seems to be supported by the following studies.
  \cite{pee02} found that the specific position of the 6.2 and 7.7
  $\mu$m PAH bands are related to each other. \citet{bre05} found that
  the centroid of the 7.7 $\mu$m feature depends on G$_0$/n$_e$, which
  determines the charge balance in reflection nebulae.

  %{\bf (I could substitute the previous 2 paragraphs by this shorter
  %  one, whatever people prefer).} There have been several hypotheses
  %to explain the changes in wavelength of the peak of the PAHs. For
  %instance, N-substituted PAHs (PANHs; Peeters et al.  2002, Hudgins
  %et al. 2005) and the ratio of $^{13}$C/$^{12}$C have also been
  %proposed to explain the observed variations in the 6.2~$\mu$m PAH
  %band \citep{wad03}.  However, none of these explanations are
  %completely satisfactory yet.

  Another mechanism can explain the shift in the PAH bands.  There is
  increased observational evidence \citep{got00, slo07} that the
  change in profile is in agreement with hydrocarbon mixtures that
  contain both aromatic and aliphatic bonds (see \S3.3). In this
  scenario, in class C spectra the hydrocarbons are a relatively
  unprocessed mixture of aliphatics and aromatics, and as the
  radiation field becomes stronger the aliphatic bonds break to expose
  the aromatic bonds and give rise to class B and eventually class A
  spectra.  This hypothesis has recently gained support from
  laboratory measurements of the 6.2~$\mu$m PAH feature by
  \citet{pin08} using carbonaceous soot. While they acknowledge that
  the astro-PAHs still need the right laboratory analogues, they have
  been able to reproduce the 6.3$\mu$m peak using aliphatic material
  (class C) and the 6.2$\mu$m PAH (class A) with more mature PAHs.
  This line of work is very promising.

  We have seen that the variations in peak position of the PAHs in the
  MCs are similar to those of their Galactic counterparts. The PAH
  profiles do not seem to vary significantly from Galactic PNe either.
  It seems that the overall conditions that produce the variations in
  the features in PNe are similar in the MW and in the MCs.

  \subsection{PAH Clusters}

  In addition to showing strong PAHs and having similar spectra,
  SMP~LMC~36, SMP~LMC~38, and SMP~LMC~61 all have a plateau between
  10-14~$\mu$m. These are the same three objects which do not show SiC
  but do show MgS (see \S4 and \S5 respectively). This plateau has not
  been identified as any PAH overtone or combination of bands
  \citep{ker00}, but it has been suggested to be related to PAH
  clusters \citep{bre89,all89,bus93,rap05,job08}.

  \subsection{Hydrogenated Amorphous Carbon}

  The spectra of SMP~LMC~02 and SMP~SMC~24 show unusual emission of
  hydrocarbons between 6 and 9~$\mu$m, and the spectrum of SMP~SMC~24
  also shows smaller bumps at 17.5~$\mu$m and 19.5~$\mu$m (see
  Figure~1). The broad bump from 6-9$\mu$m is similar (although it
  differs in detail) to that shown by the PPN IRAS~22272+5435.
  \citet{bus93} identified this emission with large hydrocarbons, e.g.
  PAH clusters and Hydrogenated Amorphous Carbon (HAC). HAC has been
  observed at 3.4~$\mu$m in IRAS~22272+5435 by \citet{got03}, and it
  is therefore possible that the 6-9~$\mu$m emission in these nebulae
  is due to HAC\footnote{We identify the emission in SMP~SMC~24 from
    9-14~$\mu$m with SiC, but the feature also resembles that of
    IRAS~22272+5435. If HAC is responsible for the emission in the
    6-9$\mu$m region, it is possible that they also contribute to the
    region where SiC is emitting. However, the SiC EW measurement
    falls well within in the relation shown in Figure~4, and its
    profile in Figure 5 resembles other SiC profiles.}.  Aliphatics
  are also seen in the IRS spectrum of the pre-planetary nebula
  SMP~LMC~11 \citep{ber06}, which shows many molecular absorption
  bands that are the building blocks from which more complex
  hydrocarbons are produced. In addition, \citet{kra06} detected
  C$_2$H$_2$ (with aliphatic bonds) as well as class C PAHs in the
  post-AGB star MSX~SMC29. \citet{got03} and \citet{slo07} propose
  that carbonaceous materials are synthesized as large HAC compounds,
  and that the aliphatic bonds are broken as they get exposed to
  harder radiation fields, eventually leaving only aromatic bonds
  (those of PAHs).  \citet{job08} find that the broad feature at
  8.2$\mu$m, which is associated with HAC, is more enhanced in young
  PNe. In this view, the possible presence of HAC in SMP~SMC~24 and
  SMP~LMC~02 would indicate that these PNe are very young. This is
  strengthened by the fact that only weak fine-structure lines of very
  low ionization potential are seen in the spectra of both objects.

\section{SILICON CARBIDE}

Figure~1 shows that most of the carbon-rich PNe have a broad feature
from $\sim$9 to 13~$\mu$m which peaks between 10.8 and 11.7~$\mu$m. We
identify this feature as SiC. This feature is very prominent and
explains the high IRAS flux (F$_{12}$/F$_{25}$) derived by
\citet{zij94} for some of these sources.

The SiC band is rare in Galactic PNe; from the available {\em ISO}-SWS
PNe sample only IC~418 shows the feature, and it is also present in
M1-20 from UKIRT spectra \citep{cas01}. However, the SiC band is
commonly seen in Galactic AGB stars.  \citet{sta07} propose an
evolutionary sequence, where the SiC is seen in young PNe where PAHs
are not yet formed. They make this claim based on the weak PAHs that
are seen when the SiC is present in their sample. In our sample we see
no such relation, with several sources showing strong PAH and strong
SiC emission.  Besides, if the suggestion by \citet{sta07} was correct
it should also hold for the MW, but hardly any MW PNe show SiC, while
the PAHs are similar in the MW and MC. Thus, there should be another
reason related to the environment that influences SiC formation and
emission.  As in the sample of \citet{sta07}, the PNe in our sample
with SiC also have the smallest diameters\footnote{The Galactic PNe
  IC~418 and M1-20 also have small sizes.}, hinting at an evolutionary
process, but also maybe suggesting to a slower expansion velocity. At lower
metallicities the mass-loss may be slower because there is less dust
to drive the outflow. On the other hand, since the carbon in the dust
is produced within the star, one may not expect a decreased
(carbon-dust) mass-loss rate with metallicity.  \citet{mat07} and
\citet{slo08} did not find a relation between metallicity and
mass-loss in the sample of AGB stars in the Fornax galaxy that they
studied. The SiC is inherited from the AGB phase. Likely, at lower
metallicity, the transition from O-rich to C-rich occurs earlier in
the evolution on the AGB, and hence much of the Si is channeled to a
carbonaceous condensation (SiC) rather than an oxide (e.g.,
silicates).

We measured the SiC feature by integrating the continuum subtracted
spectra from 9 to 13.2~$\mu$m, and removing the flux of the
11.2~$\mu$m PAH feature and the \ion{[Ne}{2]} line when present
(Table~2). Our data suggest that the feature is highly sensitive to
the radiation field (at least at the metallicities we are probing).
While SiC is present in PNe with strong \ion{[Ne}{3]} lines (41.0~eV),
the feature is absent when either the \ion{[O}{4]} (54.9~eV) or the
\ion{[Ne}{5]} (97.1~eV) lines are detected. Thus it seems that photons
with energies of about 55~eV destroy SiC grains.  Figure~4 plots the
equivalent width\footnote{The equivalent width is plotted instead of
  the integrated flux because it is independent of distance.} of the
SiC feature against the ratio of the \ion{[Ne}{3]} (15.56~$\mu$m) to
\ion{[Ne}{2]} (12.81 $\mu$m) line fluxes measured by \citet{ber08}
which traces the hardness of the radiation field.  The sample is
unfortunately small but there is a clear tendency of lower SiC
equivalent width with increasing radiation field hardness.  IC~418
also falls on this relation. The same tendency is seen if the
\ion{[S}{4]} (10.51~$\mu$m) to \ion{[S}{3]} (18.71~$\mu$m) line ratio
is used.  This corroborates the relation of the SiC with ionization.

\begin{figure}
  \begin{center}
    \includegraphics[width=8cm]{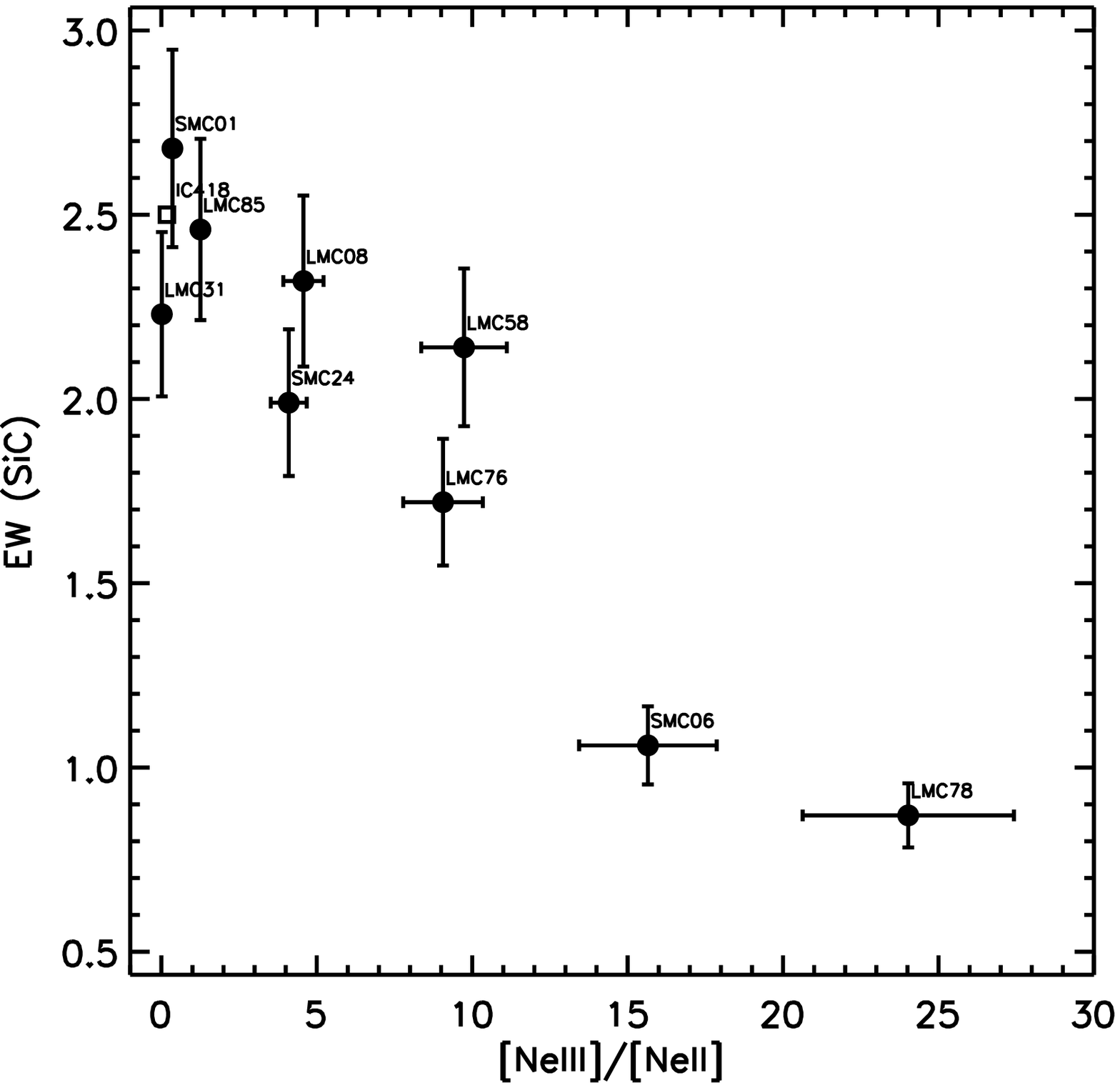}
  \end{center}
  \caption{The equivalent width (in $\mu$m) of the SiC decreases with
    the hardness of the radiation field as given by the
    \ion{[Ne}{3]}/\ion{[Ne}{2]} ratio. The Galactic PN IC~418 is shown as
    an open square. \label{sicline_f}}
\end{figure}

Figure~5 shows the profile of the SiC feature.  The 11.2~$\mu$m PAH
feature has been removed, and the positions of the fine-structure
lines of \ion{[S}{4]} and \ion{[Ne}{2]} are indicated by vertical
lines.  It is clear that the profile varies from source to source.
This is interesting because in Galactic and MC AGB stars the feature
is very constant in shape. For comparison the red SiC profile of the
Galactic PN IC\,418\footnote{For M~1-20 only 8-13~$\mu$m spectrum is
  available and no \ion{[Ne}{3]} line could be measured.} and the blue
profile of a post-AGB object which encompass most of our sources
(although we have some bluer ones) are also shown. Several factors
could be responsible for shaping the feature: shape, size, extinction,
and structure of the grains.

\citet{gil72} and \citet{tre74} first suggested that a distribution of
shapes could produce the observed emission between longitudinal
(10.3$\mu$m) and transverse (12.6$\mu$m) optical phonon modes.
According to \citet{spe05}, SiC at $\sim$10.85~$\mu$m can be
reproduced with larger spherical grains, where the longitudinal mode
becomes more significant. \citet{cle03} give an alternative
explanation where the shift in position is attributed to different
degrees of agglomeration of $\beta$-SiC nano-particles (with
agglomerated SiC shifting the feature to the red compared to
non-agglomerated material).  The change in position in our sample may
reflect aggregation or changes in size, but it is also possible that a
change in grain shape could produce the observed shift.  We note that
extinction is not likely to cause the variation of the SiC profile in
our sample.  SiC represents only a fraction of the dust, even in
optically thick shells \citep[e.g.][]{gri90,slo95}.  Those rare cases
where SiC is seen in absorption may be associated with a very
short-lived phase on the AGB \citep{spe09}. By the PN phase, the dust
shell would have dissipated and once again be optically thin.  The
extinction in the sources is in fact very low with E$_{B-V}$ ranging
from 0.04 to 0.36 \citep{ber08}.

\begin{figure}
  \begin{center}
    \includegraphics[width=9cm]{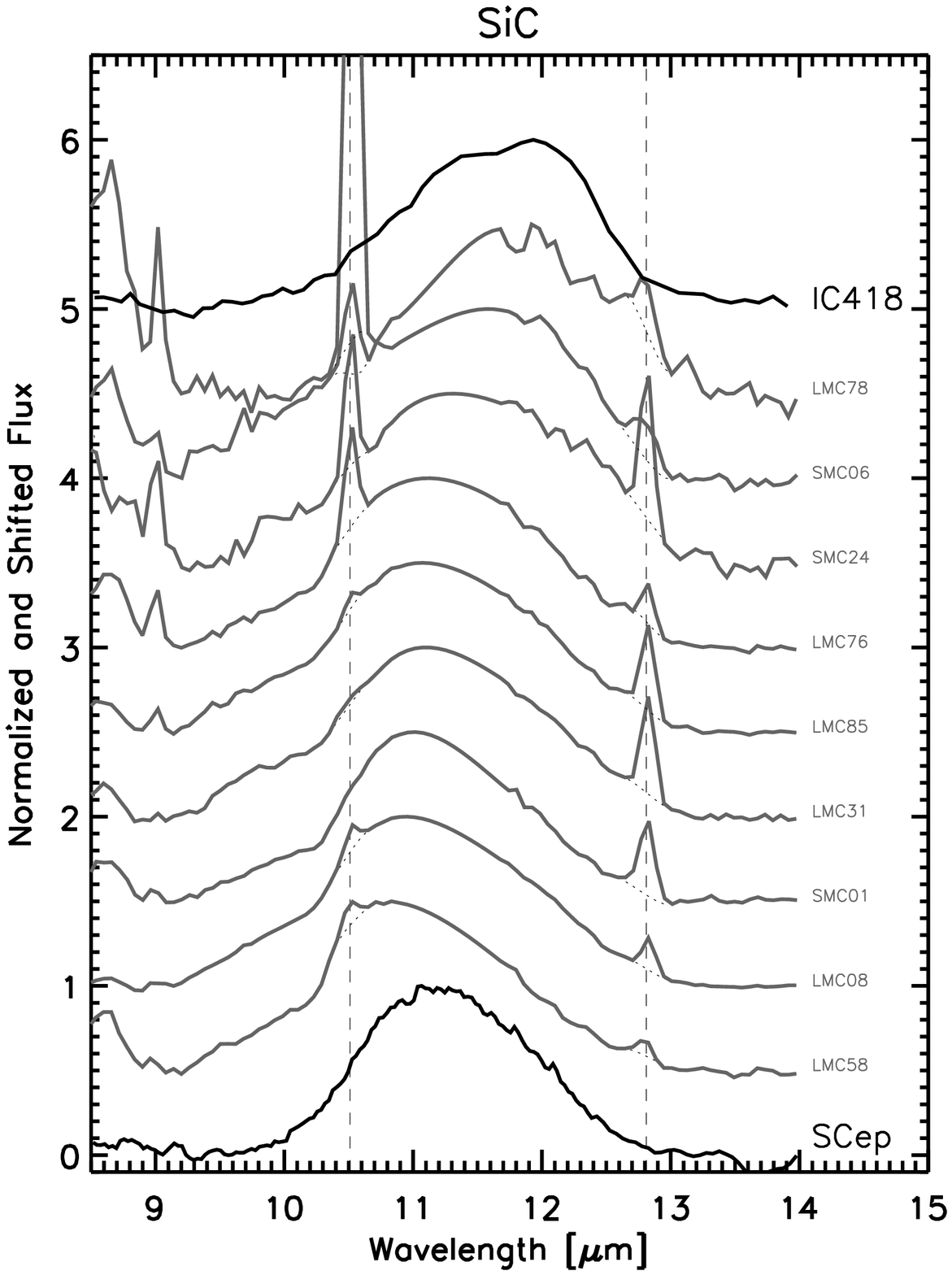}
  \end{center}
  \caption{SiC feature in the MC PNe (in grey). The positions of the
    \ion{[S}{4]} and \ion{[Ne}{2]} lines are indicated with vertical
    lines with an underlying local continuum. For comparison, the SiC
    profile of the carbon star SCep and the PN IC~418 (in black) are
    also displayed.\label{sicprofile_f}}
\end{figure}

Comparisons between samples of Galactic AGB stars with other
metal-poor carbon stars in the LMC \citep{zij06}, SMC \citep{lag07},
and the Fornax dwarf spheroidal \citep{mat07} show that the strength
of the SiC feature decreases at lower metallicity.  These studies
argue that this is driven by the lower silicon abundance at lower
metallicity.  The observational evidence from the PNe seems to
indicate that the silicon abundance is not the main driver of the
feature (at least down to the SMC metallicity). The reason the feature
is stronger in PNe than in AGB stars may be related to
the higher temperature of the central star in the PN phase.  In order
to be sufficiently heated, SiC needs to be close to the central star
(SiC absorbs shortward of 0.4~$\mu$m). Its presence in a disk may
provide the necessary conditions for its formation. This incidentally
could allow the grains to grow larger and could also affect its shape.
Small grains cannot be ruled out. The strong SiC features in our
spectra may be the result of a significant population of smaller
grains instead of more SiC (by mass). This is consistent with the
picture proposed by \citet{spe05} where the grains get smaller as the
star evolves although it does not explain why it is not seen in
Galactic PNe.

The condensation sequence of SiC is still a matter of debate. In their
study of carbon stars, \citet{lag07} suggested that, based on the
condensation temperatures of graphite and SiC, the SiC is deposited on
carbon grains in the MC, while in the MW it is the other way around.
Our observations agree with this sequence.  \citet{lei08}, in their
IRS study of 19 AGB stars in the LMC, propose a more complex carbon
condensation sequence which involves amorphous carbon, SiC, and MgS,
based on the strength of these features and the mass-loss rate.  For
the MCs they favor a sequence SiC$\rightarrow$(C,SiC)$\rightarrow$MgS
(with MgS coating a layer of C and SiC). In this view MgS needs a
layer of SiC to be formed.  In our sample, only SMP~LMC~36,
SMP~LMC~38, and SMP~LMC~61 show MgS but no SiC, and while the strength
of the MgS feature has not been calculated it is clear that these
three PNe present a very weak feature as compared to the rest of the
sample (e.g. SMP~LMC~08, SMP~LMC~31).  Furthermore in this coating
sequence proposed by \citet{lei08}, MgS always ends up as the top
outer layer. In this scenario MgS cannot be hidden, and thus the
amount of sulfur depleted onto this feature can not be used to explain
the low sulfur abundance seen in some PNe (see next section).

\section{THE 30$\mu$m FEATURE}

The `30$\mu$m' feature has a long history, and it is common in the
spectra of carbon stars and PNe. It usually extends from 25 to
45~$\mu$m, and was first discovered by \citet{for81}.  In the early
stages it was tentatively identified as MgS by \citet{goe85}, but
several other identifications were proposed shortly afterwards
\citep[e.g.][]{dul00,pap00,gri01}.  \citet{hon02} compiled a large
number of ISO spectra of carbon-stars and PNe and modeled the feature
using MgS and a continuous distribution of ellipsoids with a grain
temperature different from that of the bulk of the dust. In the same
year, \cite{vol02} found that the feature could be resolved into two
components, a narrow 26$\mu$m and wider 33~$\mu$m feature in the SWS
data, thus casting some doubts on MgS as the carrier of the feature.
It has also been proposed that this structure at 26~$\mu$m is an
artifact produced by the SWS overlapping region between bands 3D and
3E \citep{slo03} which is not very well calibrated. These two
components can, however, be explained by differences in the shape
distribution of MgS grains, where a stronger peak near 26~$\mu$m
indicates more spherically shaped grains (S. Hony, private
communication).  Within this frame one could speak of two MgS
components; one more spherical and one less so.  Despite the caveats,
MgS is still the strongest candidate for this feature and we shall
assume this for the discussion in the rest of the paper.

\begin{figure}
  \begin{center}
    \includegraphics[width=9cm]{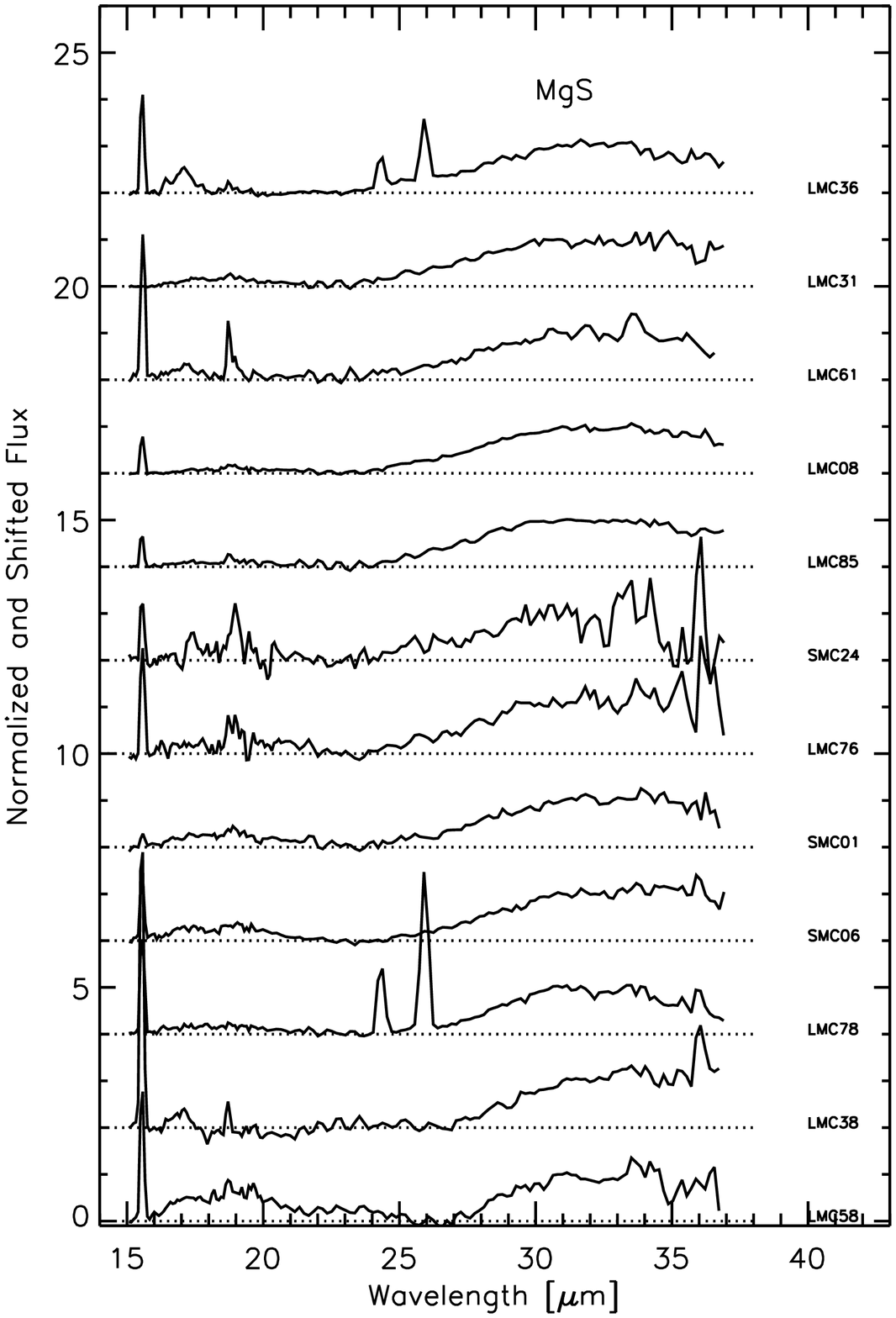}
  \end{center}
  \caption{Continuum subtracted spectra (see \S5) of the 15-37$\mu$m
    region to show the MgS feature. The feature at 16-21$\mu$m is
    discussed in \S7 and Figure~7.\label{bump1520_f}}
\end{figure}

The feature is shown in Figure~6. Unfortunately, due to the wavelength
coverage, the IRS modules miss the long-wavelength end of the feature,
and the spectra also get noisy at long wavelengths. For this reason we
have not attempted to measure the strength of the feature because the
measured flux would be unreliable.  Much care was taken defining the
underlying continuum, but the reader should bear in mind that while
Figure~6 shows the PNe with MgS, the figure does not necessarily
reflect the true strength of the feature.  MgS is present in 12 of the
PNe, all of the PNe showing PAHs except SMP~LMC~36, SMP~LMC~38 and
SMP~LMC~61, and in SMP~SMC~24 where HAC may be present (\S3.3). This
feature can carry up to 30\% of the infrared luminosity in the
Galactic PNe and carbon stars \citep{hon02}.  Therefore, if the
feature is due to MgS a significant amount of magnesium and sulfur
could be tied up in MgS. It is known that Galactic PNe show a sulfur
depletion when compared to Solar \citep{mar,pot06}. The sulfur
abundances of six of the PNe in this sample are also lower than those
found in MC H\,II regions \citep{ber08}.  This sulfur under-abundance
has even been seen in studies of H\,II regions in Local Group galaxies
where it has also been suggested that maybe sulfur is depleted onto
dust in these systems \citep{rub07}. We tried to correlate the
presence of this feature with the sulfur abundance but found no clear
relation.  Some PNe with MgS do show a low sulfur abundance but other
PNe with MgS show normal sulfur abundances.  This argues against
sulfur being significantly depleted onto this feature (or against MgS
being the carrier). There may be other sulfur-based features (e.g.
FeS) on which sulfur is depleted, but none have been detected in the
spectra.

\section{SILICATES}

The most common form of silicates are amorphous silicates, with
features at 9.8~$\mu$m (Si-O stretching mode) and 18.5~$\mu$m (O-Si-O
bending mode). Only two of the 25 objects in our sample show silicate
emission, SMP~LMC~53 and SMP~LMC~62 (see Fig.~1).  SMP~LMC~53 does not
show the 9.8~$\mu$m stretching mode. This is normal and there are not
many Galactic PNe with this feature either (e.g. NGC6302, NGC6543,
Mz3).  This is because as the nebula expands, silicate dust grains
cool and are less efficient at emitting at 10~$\mu$m.  There are no
signs of the presence of crystalline silicates in our sample, even in
the high-resolution spectra.  In the sample of \citet{sta07} there are
three additional oxygen-rich PNe (two in the LMC and one in the SMC).
Adding them to our sample results in four oxygen PNe in the LMC out of
44, and one in the SMC out of 23. In the Galactic sample the ratio of
PNe showing oxygen-rich dust (silicates) compared to carbon-rich dust
(PAHs) is higher than in the MCs. This has been found before in the
study of AGB stars by \citet{mat07} where they found that the number
of carbon-rich stars in the MCs was larger than expected. The reason
for this is the decrease of the mass-limit ($\sim$1.5~M$_\odot$ for
the MW) at which the third dredge-up occurs at lower metallicities
\citep{mar}. This dredge-up essentially mixes large amounts of $^4$He
and $^{12}$C during thermal pulses and turns the envelope of the star
carbon-rich (C/O$>$1). In addition the original lower oxygen content
means that less carbon is needed to turn the C/O ratio larger than 1
\citep{slo08}. Therefore we expect a larger number of PNe with
carbon-rich chemistry than oxygen-rich ones at lower metallicity.

SMP~LMC~28 shows a feature around 33$\mu$m. This feature could be
attributed to crystalline silicates but other expected crystalline
features at 28 and 23~$\mu$m are not detected. Lowering the
temperature of the crystalline silicates (less than 100~K) weakens the
23 and especially the 28~$\mu$m features, but given the caveats, we
leave this identification as tentative.

\section{OTHER FEATURES}

Many of the PNe display some kind of structure from around
$\sim$16-22~$\mu$m (Fig.~7). This feature could simply be the result
of dust peaking at 20~$\mu$m giving the spectrum a peculiar shape, but
black- and grey-body fits (although uncertain to some degree) cannot
match the region which should be dominated by emission of very small
grains.  If the peculiar shape is in fact a feature, the
identification is not clear.  Some features emitting around these
wavelengths include: 1) the C-C-C PAH bending mode which produces a
plateau from 15 to 20~$\mu$m, 2) amorphous silicates (O-Si-O
stretching mode), and 3) the still unidentified 21~$\mu$m feature
\citep{hri00,hon01}.  The feature is difficult to isolate because of
the many other features present in the spectra (PAHs, plateaus, SiC,
MgS) which make it difficult to determine a local continuum. However,
we fitted a grey-body function \citep{hon02} to the spectrum and
compared with similarly fitted spectra of a YSO with a PAH plateau
(S106 IRS) and two PNe with amorphous silicates from our own sample
(SMP~LMC53 and SMP~LMC62).  Figure~7 shows this comparison. The PAH
plateau is clearly different from the feature and can be ruled out.
The feature resembles amorphous silicates, and although the comparison
is still not convincing, the reader should bear in mind the difficulty
in assessing the continuum.  If it were silicates it would then mean
that a large fraction of the PNe presented here show a dual chemistry
(PAHs and silicates). The 21$\mu$m feature was first identified in
PPNe by \citet{hri00} and was later detected in some PNe by
\citet{hon01}.  While not shown, we have compared the bump with {\em
  ISO}-SWS spectra of evolved objects showing the 21$\mu$m feature,
and the bump is distinctly very different in profile and position,
ruling this identification out.

\begin{figure}
  \begin{center}
    \includegraphics[width=9cm]{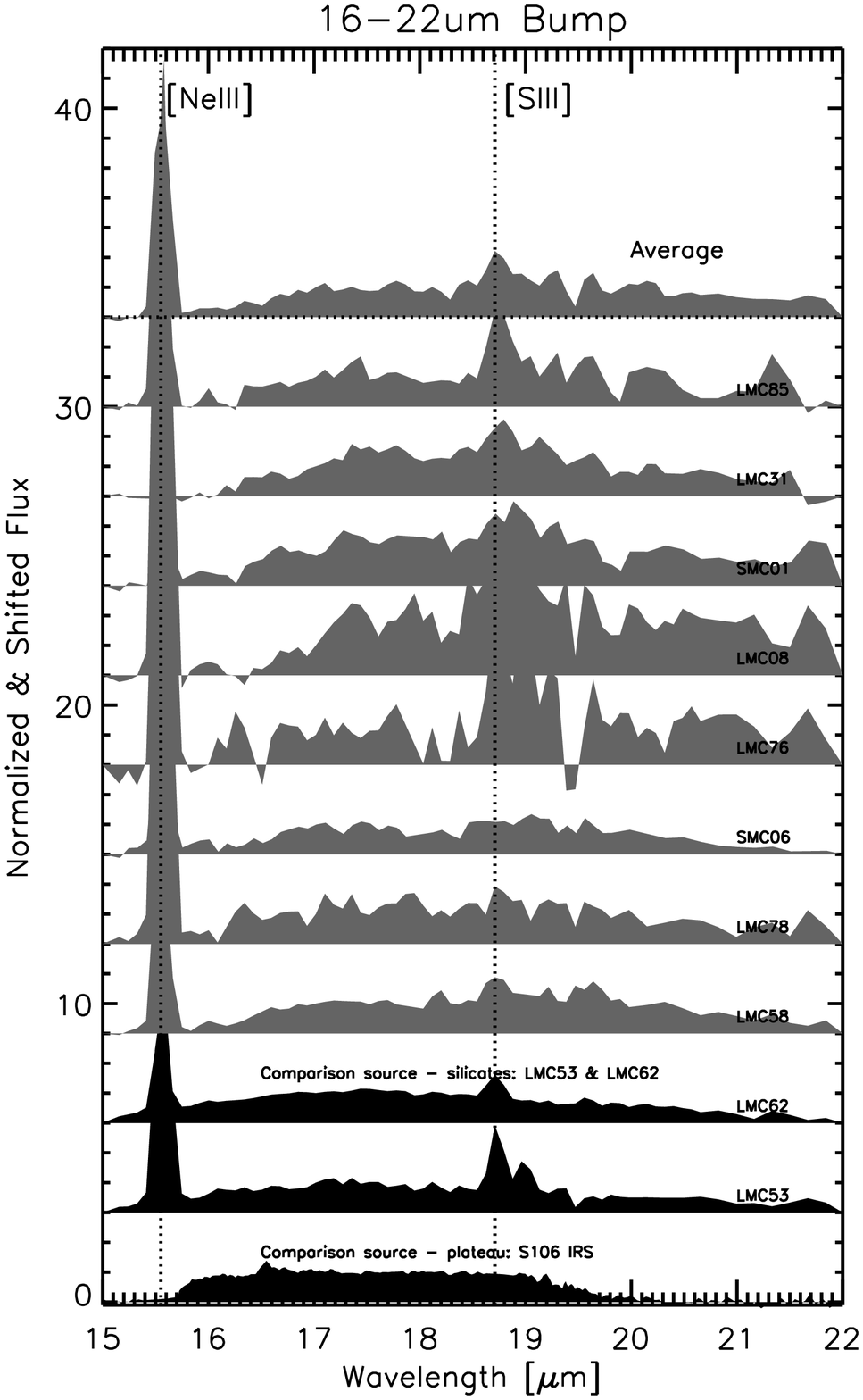}
  \end{center}
  \caption{Continuum subtracted spectra (see \S6) of the 15-22~$\mu$m
    region. The `bump' in the sources (grey) is compared to two PNe
    with amorphous silicates and a YSO with a typical PAH plateau
    emission (black).\label{bump1520_f}}
\end{figure}

In SMP~SMC~11 the continuum is very peculiar, showing a bell-like shape
starting around 15~$\mu$m. Metal-oxides could be a possible
constituent to produce this excess emission (B. Sargent \& W.  Forrest
2008 private communication). Although the spectrum is noisier, the
continuum in SMP~SMC~28 is similar to that of SMP~SMC~11 with the dust
continuum steepening very suddenly after 12~$\mu$m. The peak of the dust
continuum of these two sources lies between 30 and 35~$\mu$m,
indicating cold dust with respect to the rest of the sample.

\section{Summary and Conclusions}

%Some objects show less common features in their spectra. 

We have presented low-resolution spectra taken with the {\em Spitzer
  Space Telescope} of 25 planetary nebulae in the Magellanic Clouds
(18 in the LMC and 7 in the SMC). This is the same sample of objects
examined at higher resolution by \citet{ber08}, but now we focus on
the analysis of the dust features. The spectra show the typical
features seen in Galactic PNe, but additionally some objects show very
unusual dust.

Out of the 25 PNe, 14 show dust features characteristic of carbon-rich
dust such as amorphous carbon and PAHs. Only two objects (SMP~LMC~53
and SMP~LMC~62) display oxygen-rich dust (amorphous silicates), and
none show dual chemistry.  The high proportion of nebulae with
carbon-rich dust in the MCs compared to Galactic PNe is expected and
reflects the efficiency of the third dredge-up at lower metallicities.

The ratio of ionized versus neutral PAHs in the MCs and MW correlate
well, with a slope of 1.3. This slope agrees with what is found for
H\,II regions. However, while H\,II regions are segregated in
metallicity (higher ionization with higher metallicity), the PNe
spread all over the range of ratios. The profiles of the PAHs are also
similar to their Galactic counterparts.

The similarity seen in the PAH bands contrasts with the strong SiC
feature which is seen in nine of the MC PNe.  This feature is
commonly seen in AGB stars, but it is very rare in Galactic PNe (only
2 in the Galactic sample). All of these objects show PAHs as well.
Several studies in the literature comparing AGB stars in the Galaxy,
LMC, SMC, and the metal-poor Fornax Dwarf spheroidal, observed that
the SiC feature decreases at lower metallicities, and this makes our
finding even more remarkable. While at lower metallicities one expects
less silicon to be available, it may be that the lower critical
abundance of silicon to significantly suppress the SiC feature is not
reached at the metallicities of the LMC and SMC.  The equivalent width
of the feature decreases with the hardness of the radiation field, and
it is not seen in high excitation PNe, hinting at the destruction of the
feature in hard radiation fields or preventing its formation.

The `30$\mu$m' feature usually attributed to MgS is seen in 12 nebulae,
but because of the wavelength range coverage of the IRS we are unable
to reliably measure its flux. In recent studies it has been suggested
that sulfur may be depleted onto dust to explain the anomalous gas
phase abundance of this element in many PNe and H\,II regions.
However, we do not find a clear correlation of PN with MgS having a
lower abundance of sulfur in the sample.

There are three PNe that show MgS but no SiC. Instead of SiC the
spectra of these PNe show strong PAHs with a plateau from 10-15$\mu$m.
HAC may be present in SMP~LMC~02 and SMP~SMC~24, which would imply that
these are young PNe. This is supported by the weak and low excitation
lines in the spectra of these objects.

%A broad band from 16 to
%22~$\mu$m is seen in a handful of objects. The origin of this feature
%is not entirely clear but it resembles amorphous silicates.  If it
%were silicates it would mean that a high fraction of the objects in
%our sample shows dual chemistry.

The IRS spectrograph on board the {\em Spitzer Space Telescope} has
shown that the dust in MC PNe is both similar to and different from
that of Galactic PNe. This is important for understanding the
influence of metallicity on the different dust features that are
present in the mid-IR wavelength range. Future observatories such as
{\em Herschel} and {\em JWST} will be great aids  in complementing the
on-going work of {\em Spitzer}.  {\em Herschel} will
allow the extension of the mid-IR region probed by {\em Spitzer} to much
longer wavelengths, and {\em JWST} will make it possible to extend the
study of PNe to the furthest galaxies in the Local Group.

\acknowledgments We would like to thank B. Forrest and B. Sargent for
insightful comments and discussions. This work is based on
observations made with the {\em Spitzer Space Telescope}, which is
operated by the Jet Propulsion Laboratory, California Institute of
Technology under NASA contract 1407. Support for this work was
provided by NASA through Contract Number 1257184 issued by
JPL/Caltech.

\end{document}